\documentclass[twocolumn]{pasj00}
\SetRunningHead{Astronomical Society of Japan}{A new cluster of galaxies, Suzaku J1759$-$3450}

\usepackage{times}
\usepackage{color}
\usepackage{ulem}

\begin{document}

\title{A New Cluster of Galaxies towards the Galactic Bulge, Suzaku J1759$-$3450}
\author{Hideyuki \textsc{Mori},\altaffilmark{1}
Yoshitomo \textsc{Maeda},\altaffilmark{2} 
Akihiro \textsc{Furuzawa},\altaffilmark{3} 
Yoshito \textsc{Haba},\altaffilmark{4}
and Yoshihiro \textsc{Ueda},\altaffilmark{5}}
\altaffiltext{1}{Kobayashi-Maskawa Institute for the Origin of
Particles and the Universe, Nagoya University, Furo-cho, Chikusa-ku,
Nagoya, 464-8602}
\altaffiltext{2}{Department of Space Astronomy and Astrophysics,
Institute of Space and Astronautical Science (ISAS), Japan Aerospace
Exploration Agency (JAXA), 3-1-1, Yoshinodai, Chuo-ku, Sagamihara,
252-5210}
\altaffiltext{3}{Institute of Liberal Arts and Sciences, Nagoya
University, Furo-cho, Chikusa-ku, Nagoya, 464-8602}
\altaffiltext{4}{Aichi University of Education, 1, Hirosawa,
Igaya-cho, Kariya, 448-8542}
\altaffiltext{5}{Department of Astronomy, Graduate School of Science,
Kyoto University, Sakyo-ku, Kyoto, 606-8502}
\email{mori@u.phys.nagoya-u.ac.jp}

\maketitle

\begin{abstract}
We observed an extended X-ray source designated as Suzaku J1759$-$3450
with the Suzaku and Chandra observations towards 1RXS
J175911.0$-$344921, which is an unidentified X-ray source listed in
the ROSAT Bright Source Catalogue.  A conspicuous emission line at
$6$~keV was also found in the Suzaku J1759$-$3450 spectrum.  Assuming
the emission line to be K emission from He-like Fe ions, we inferred
Suzaku J1759$-$3450 to be an extragalactic object located at $z =
0.13$.  The radial profile of the surface brightness in the
$0.5$--$10$~keV band was explained well with an isothermal
$\beta$-model of $r_{\rm c} = \timeform{1.61'}$ and $\beta = 0.78$.
The X-ray spectrum was well reproduced by an optically-thin thermal
plasma with the electron temperature of $kT_{\rm e} = 5.3$~keV
attenuated by the photoelectric absorption of $N_{\rm H} = 2.3 \times
10^{21}$~cm$^{-2}$.  The bolometric X-ray luminosity of $L_{\rm X} (r
< r_{500}) = 4.3 \times 10^{44}$~erg~s$^{-1}$ is consistent with that
expected from the $L_{\rm X}$--$T$ relation of clusters of galaxies.
In terms of the spatial extent, the X-ray spectrum, and the bolometric
luminosity of the X-ray emitting gas, we concluded that Suzaku
J1759$-$3450 is a new cluster of galaxies.
\end{abstract}

\section{Introduction}
\label{section:intro}
Population studies of X-ray sources are vital for our understanding of
the dynamical and chemical evolution of the host galaxies since the
luminous X-ray sources reflect the current activities or endpoints of
the high-mass stars.  Many long-exposure observations were dedicated
to elucidating the X-ray population in our Galaxy such as ASCA
Galactic Plane Survey \citep{2001ApJS..134...77S}, XMM-Newton Galactic
Plane Survey \citep{2010A&A...523A..92M}, and ChaMPlane
\citep{2005ApJ...635..920G}.  These extensive observing campaigns had
the advantage of detecting extremely faint X-ray sources, which has
never been achieved for the extragalactic X-ray surveys.
Nevertheless, the sky coverage for the Galaxy is limited due to the
deep pointing observations.

Hence, the more effective approach is probably required to decipher
the discrete X-ray population of the Galactic components -- Galactic
center, disk, and bulge.  Because of the most recent survey which had
completed a full sky coverage with an imaging X-ray optics, the ROSAT
All-Sky Survey (RASS) provides the useful database to study the X-ray
population with the flux limit down to $\sim
10^{-12}$~erg~s$^{-1}$~cm$^{-2}$.  The RASS detected many unidentified
X-ray sources towards the Galactic bulge where the interstellar medium
in the line of sight does not hamper the soft X-rays in the ROSAT PSPC
band ($0.1$--$2.4$~keV).  This fact motivates us to carry out the
follow-up observations to investigate their nature (e.g.,
\cite{2012PASJ...64..112M}).  In the present paper, we report the
X-ray properties of the cluster of galaxies newly detected with such
an X-ray observation.

The paper is organized as follows; we first present the observations
with Suzaku and Chandra in section \ref{section:observation}.  The
X-ray data analysis is explained in section \ref{section:analysis}.
After the discussion (section \ref{section:discussion}), we give a
short summary of the source identification in section
\ref{section:summary}.  Throughout the paper, we used the cosmological
parameters of $H_{\rm 0} = 70$~km s$^{-1}$ Mpc$^{-1}$, $\Omega _{\rm
M} = 0.27$, and $\Omega _{\rm \lambda} = 0.73$, which represent the
Hubble constant, dimensionless density parameters of the matter and
dark energy, respectively.  We also note that the errors represent the
$90$\% confidence limit, unless otherwise mentioned.

\section{Observation and data reduction}
\label{section:observation}
\begin{table*}[bhtp]
\begin{center}
\caption{Observation log}
\label{table:observation}
\begin{tabular}{lcccc}
\hline
        & Obs. ID & Start time (UT) & End time (UT) & Exposure\footnotemark[$*$] \\
\hline
Chandra & 12946\footnotemark[$\dagger$] & 2011/10/13 18:22:25 & 2011/10/13 19:54:51 & 3.50 (ACIS-S3) \\
Suzaku  & 406019010 & 2012/03/07 21:40:15 & 2012/03/08 21:54:13 & 40.2 (XIS), 34.7 (HXD) \\
\hline
\multicolumn{5}{@{}l@{}}{\hbox to 0pt{\parbox{180mm} {\footnotesize
 \footnotemark[$*$] Effective exposure of the screened data. The HXD exposure is dead-time corrected.
 \par\noindent 
 \footnotemark[$\dagger$] Sequence number is 900978.
 }\hss}}
\end{tabular}
\end{center}
\end{table*}

1RXS J175911.0$-$344921 was first discovered by the RASS.  The full
information of the source is described in the RASS Bright Source
Catalogue (RBSC: \cite{1999A&A...349..389V}).  The source position was
$(RA, Dec)_{\rm J2000.0} = (\timeform{17h59m11.0s},
\timeform{-34D49'21.5"})$ with an uncertainty of $\timeform{32"}$,
which corresponds to ($l$, $b$) = $(\timeform{356.3818D},
\timeform{-5.4660D})$ in Galactic coordinates.  The count rate and the
net exposure of the source is $0.12 \pm 0.03$~counts~s$^{-1}$ and
$196$~s, respectively.  Assuming an absorbed power-law spectrum with
$\Gamma = 2$ and $N_{\rm H} = 2.3 \times 10^{21}$~cm$^{-2}$, this
count rate corresponds to the X-ray flux of $1.7 \times
10^{-12}$~erg~s$^{-1}$~cm$^{-2}$ in the $0.1$--$2.4$~keV band.  Since
the nature of the source is still unclear, we conducted the Suzaku and
Chandra observations to obtain the hard ($> 2$~keV) X-ray spectrum.

Suzaku \citep{2007PASJ...59S...1M}, the 5th Japanese-U.S. collaborated
mission, enables us to perform imaging spectroscopy in the
$0.2$--$10$~keV band with the combination of the X-ray Telescopes
(XRTs: \cite{2007PASJ...59S...9S}) and the X-ray Imaging Spectrometer
(XIS: \cite{2007PASJ...59S..23K}).  The XIS consists of four
charge-coupled devices (CCDs); three of which are front-illuminated
(FI) CCDs (XIS0, 2, 3), and the other is a back-illuminated (BI) one
(XIS1).  The XIS2 has malfunctioned since November 9, 2006, and then
we can use the two FI CCDs for the observation.  A fraction of the
imaging area of the XIS0 has also been unusable since June 23,
2009\footnote{Suzaku memo: JX-ISAS-SUZAKU-MEMO-2010-01}.  In this
observation, the XIS was operated with the normal mode without any
window options, and the spaced-row charge injection
\citep{2009PASJ...61S...9U} was applied.

Suzaku is also equipped with the non-imaging detector which covers the
hard X-ray band of $10$--$700$~keV.  This detector, called Hard X-ray
Detector (HXD: \cite{2007PASJ...59S..35T},
\cite{2007PASJ...59S..53K}), is composed of the Si-PIN photodiodes and
the scintillators utilizing the Gadolinium Silicate crystals.  The HXD
was also operated with the normal mode.  We performed the pointing
observation where the aim point of $(RA, Dec)_{\rm J2000.0} =
(\timeform{17h59m11.0s}, \timeform{-33D49'21''})$ was placed on the
center of the XIS field of view (FOV).  We list the observation log in
table~\ref{table:observation}.

We analyzed the cleaned event data provided from the Suzaku team.  The
data were processed with the standard pipeline with the version of
2.7.16.32.  The data during the passage of the South Atlantic Anomaly
(SAA) were discarded.  The data obtained with the low earth elevation
and day-earth elevation were also excised because of the increase in
the non-X-ray background (NXB).  For the XIS1, the amount of the
injected charge was increased from $2$~keV to $6$~keV in 2011 June.
Since this change also causes the increase in the NXB, we removed the
events in the 2nd trailing
rows\footnote{http://www.astro.isas.ac.jp/suzaku/analysis/xis/xis1\_ci\_6\_nxb/}.
We used the \texttt{XSELECT} ver 2.4b provided from the HEASOFT
package (version 6.12) to extract the image, light curve, and
spectrum.  The redistributed matrix file (RMF) and ancillary response
file (ARF) were made by \texttt{xisrmfgen} and \texttt{xissimarfgen}
\citep{2007PASJ...59S.113I}, respectively.  On the other hand, we made
use of the HXD background and response files provided from the HXD
team; the NXB file we used was
\texttt{ae406019010\_hxd\_pinbgd.evt.gz} (tuned;
\cite{2009PASJ...61S..17F}), and the response file was
\texttt{ae\_hxd\_pinxinome11\_20110601.rsp}.

Chandra \citep{2000SPIE.4012....2W} is the U.S. flagship mission for
X-ray astronomy.  Chandra possesses the X-ray telescopes and attitude
control system, called High Resolution Mirror Assembly (HRMA:
\cite{1997SPIE.3113...89V}), allowing to obtain X-ray images with the
unprecedented high-angular resolution.  On its focal plane, the
Advanced CCD Imaging Spectrometer (ACIS: \cite{1997ITED...44.1633B})
is placed to provide imaging spectroscopy.  The ACIS contains 10 CCD
chips; a fraction of these CCDs makes a 2$\times$2 CCD array (ACIS-I)
and the others do 1$\times$6 CCD array (ACIS-S).  The primary goal of
the Chandra observation we proposed is to determine the accurate
position of 1RXS J175911.0$-$344921.  Thus, only the ACIS-S3 chip was
turned on.  Moreover, we selected the $128$-pixel subarray mode to
avoid the pile-up effect.  The observation was performed with the very
faint mode.  The grating modules were not used.  The log of the
Chandra observation is also summarized in
table~\ref{table:observation}.

We used the level 2 event file for the analysis; the standard data
processing was applied to the data.  In order to generate the X-ray
image and spectrum described below, the Chandra Interactive Analysis
of Observations (CIAO, version 4.4:
\cite{2006SPIE.6270E..60F})\footnote{http://cxc.harvard.edu/ciao/} and
the relevant calibration files (CALDB, version 4.4.6) were utilized.

\section{Analysis}
\label{section:analysis}
\subsection{X-ray image and light curve}
\label{subsection:image_lightcurve}
\begin{figure*}[htbp]
 \begin{center}
  \FigureFile(80mm,50mm){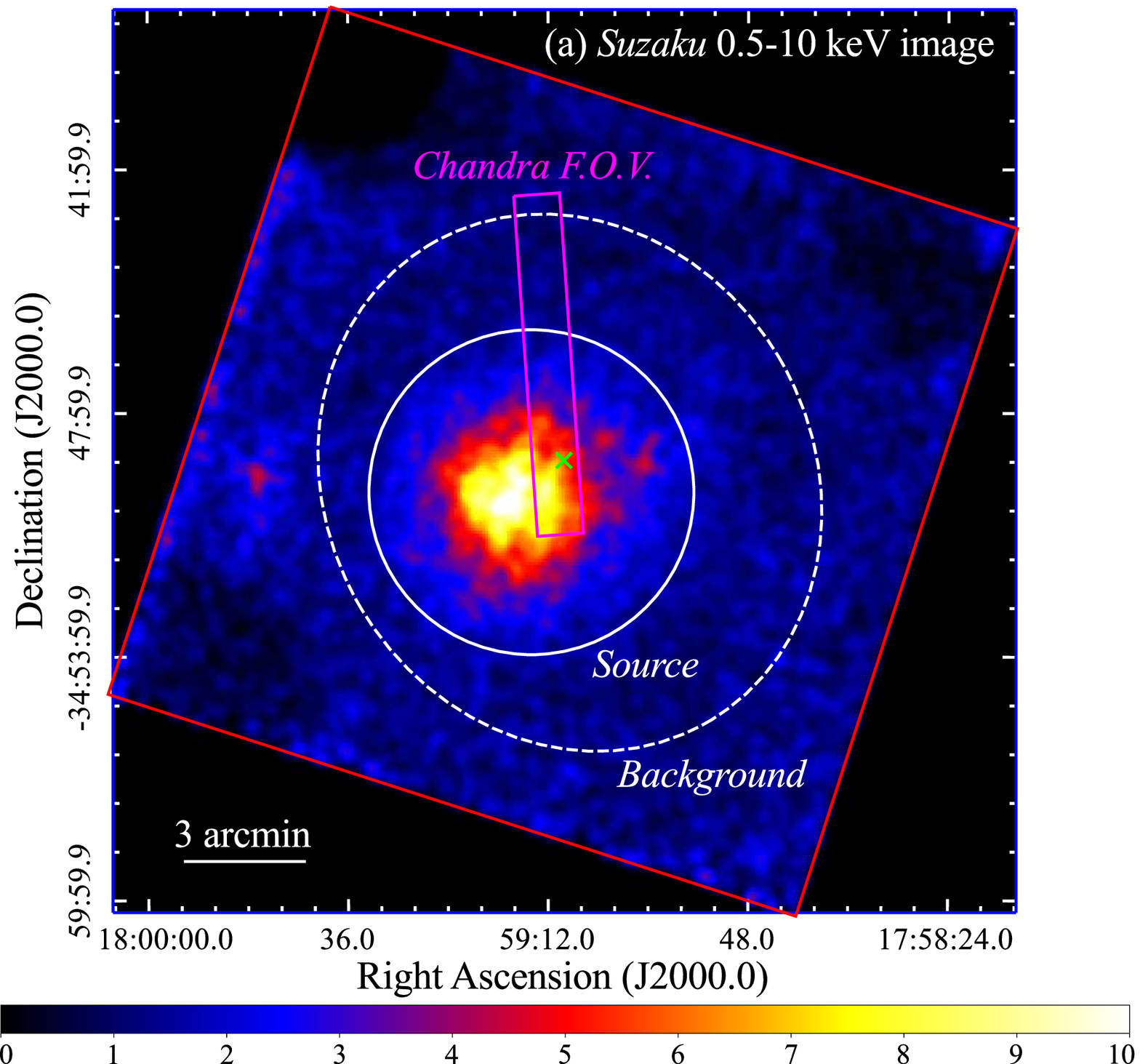}
  \FigureFile(80mm,50mm){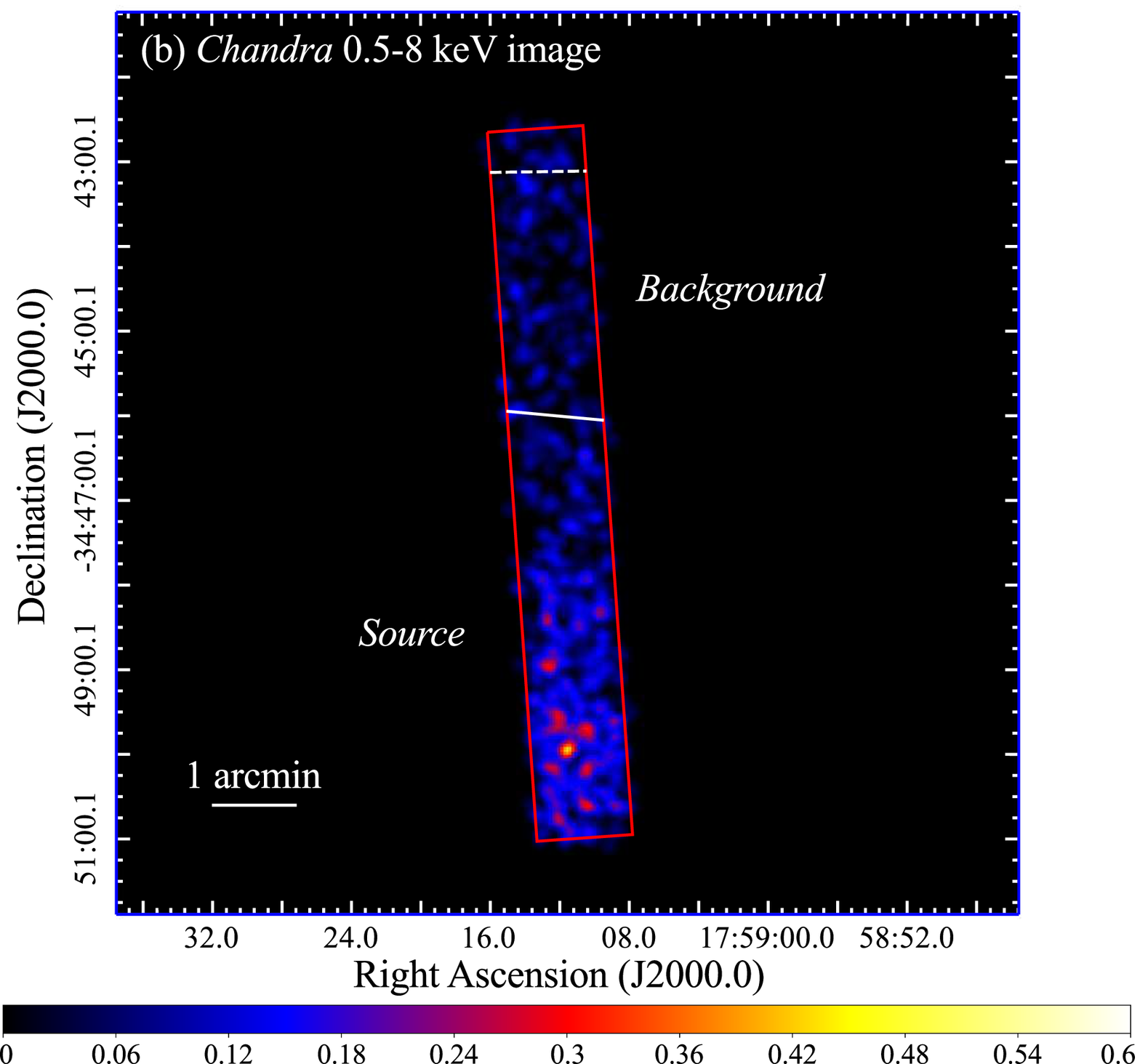}
 \end{center}
 \caption{Suzaku XIS image in the $0.5$--$10$~keV band (a) and Chandra
ACIS-S3 image in the $0.5$--$8$~keV band (b) of 1RXS
J175911.0$-$344921.  The XIS image was binned with $4 \times 4$~pixels
and then smoothed with the Gaussian kernel of $\sigma =
\timeform{0.28'}$.  The subtraction of the Non-X-ray background and
the vignetting correction were applied to the XIS image.  The source
and background regions to extract the XIS spectra are indicated with
white solid circle and dashed ellipse, respectively. The red square
and magenta rectangle represent the XIS and ACIS-S3 field of views,
respectively.  The position of 1RXS J175911.0$-$344921 is shown by a
green cross.  The ACIS-S3 was operated in the $128$-pixel subarray
mode.  We made a zoomed-up image with the binning of $4 \times
4$~pixels.  The smoothing with the Gaussian kernel of $\sigma =
\timeform{7.9"}$ was also applied.  The white solid and dashed curve
represent a part of the source and background regions in the XIS
image, which are overlapped with the ACIS-S3 field of view (red
square).  The color scales at the bottom of these panels are in unit
of photons per $\timeform{4.2"} \times \timeform{4.2"}$ (a) and
$\timeform{2.0"} \times \timeform{2.0"}$ (b) image pixel.
}
 \label{fig:Xray_images}
\end{figure*}

Figure~\ref{fig:Xray_images}a shows the XIS image in the
$0.5$--$10$~keV band of 1RXS J175911.0$-$344921.  From each CCD
sensor, we made an X-ray image binned by a factor of $4$.  After the
subtraction of the NXB image created with \texttt{xisnxbgen}
\citep{2008PASJ...60S..11T}, the vignetting correction was applied to
each image.  We estimated the vignetting effect of the Suzaku XRT for
a uniform sky with \texttt{xissim}.  These corrected images were added
together and then smoothed with a Gaussian function of $\sigma =
\timeform{0.28'}$.  We also show the X-ray image in the $0.5$--$8$~keV
band obtained with the Chandra ACIS-S3 chip in
figure~\ref{fig:Xray_images}b.  After the $4 \times 4$ pixel binning,
the smoothing with a Gaussian function of $\sigma = \timeform{7.9"}$
was applied to the image.

The XIS image clearly shows extended emission located $\sim
\timeform{1'}$ away from the position of 1RXS J175911.0$-$344921
(green cross point in figure~\ref{fig:Xray_images}a) in the south-east
direction.  The emission has a nearly circular shape with a radius of
$\sim \timeform{4'}$.  A point-like source was marginally found at
$(RA, Dec)_{\rm J2000.0} = (\timeform{17h59m47.0s},
\timeform{-34D49'28.3"})$.  The source is positionally coincident with
two radio sources, MOST 1756$-$348 and PMN J1759$-$3449, within an
uncertainty of $\timeform{15"}$.  Thus, the source may be an X-ray
counterpart of either of these radio sources.  However, the detailed
study on the source is beyond the scope of this paper.  We also
mention that there was no X-ray source at $(RA, Dec)_{\rm J2000.0} =
(\timeform{17h59m11.0s}, \timeform{-34D49'21.5"})$ in the Chandra ACIS
image.  Thus, we pay attention to this diffuse emission, designated as
Suzaku J1759$-$3450 hereafter, as an X-ray counterpart of the RBSC
source.

First, we searched for time variability of the source intensity.  For
the source-extraction region, we defined a circle with a radius of
$\timeform{4'}$ where the surface brightness decreases to $\sim 10$\%
of its peak (a white solid circle in figure~\ref{fig:Xray_images}a).
The center of the source region is at $(RA, Dec)_{\rm J2000.0} =
(\timeform{17h59m14.02s}, \timeform{-34D49'56.7"})$.  The background
region was chosen to be an ellipse excluding the source region, shown
by a white dashed ellipse in figure~\ref{fig:Xray_images}a.  The radii
of the ellipse along the major and minor axes are $\timeform{6.8'}$
and $\timeform{6'}$, respectively.  It is well known that there is
diffuse X-ray emission with characteristic Fe-K emission lines towards
the Galactic plane (Galactic Ridge X-ray Emission (GRXE):
\cite{1986PASJ...38..121K}).  Since the X-ray intensity of this
diffuse emission has a spatial dependence (e.g.,
\cite{1993ApJ...404..620Y}, \cite{2001ASPC..251..304K},
\cite{2003A&A...410..865R}), we aligned the axes of the ellipse to the
Galactic longitude and latitude to mitigate the spatial variability of
the GRXE.  The size of the ellipse was determined so as to avoid the
detector corners illuminated by the calibration sources, anomalous
rectangular region in XIS0, and the eastern point source.  We made the
background-subtracted XIS light curve for each CCD sensor, and then
fitted the light curves with a constant model.  All the fits were
acceptable with $\chi ^{2}_{\nu} = 1.1$--$1.3$.  The average count
rates were $0.153 \pm 0.004$~cts~s$^{-1}$ for XIS0, $0.188 \pm
0.005$~cts~s$^{-1}$ for XIS1, and $0.173 \pm 0.004$~cts~s$^{-1}$ for
XIS3.

We also made a hard X-ray light curve from the HXD-PIN data.  The
deadtime-corrected count rate in the $18$--$40$~keV band was $0.140
\pm 0.004$~cts~s$^{-1}$, while the NXB count rate derived from the
``tuned'' background file was $0.142 \pm 0.001$~cts~s$^{-1}$.  This
indicates that there is no significant X-ray emission above $10$~keV
within the HXD-PIN FOV.  Hence, we ignored the HXD-PIN data in the
spectral analysis.

\begin{figure}[htbp]
 \begin{center}
  \FigureFile(80mm,50mm){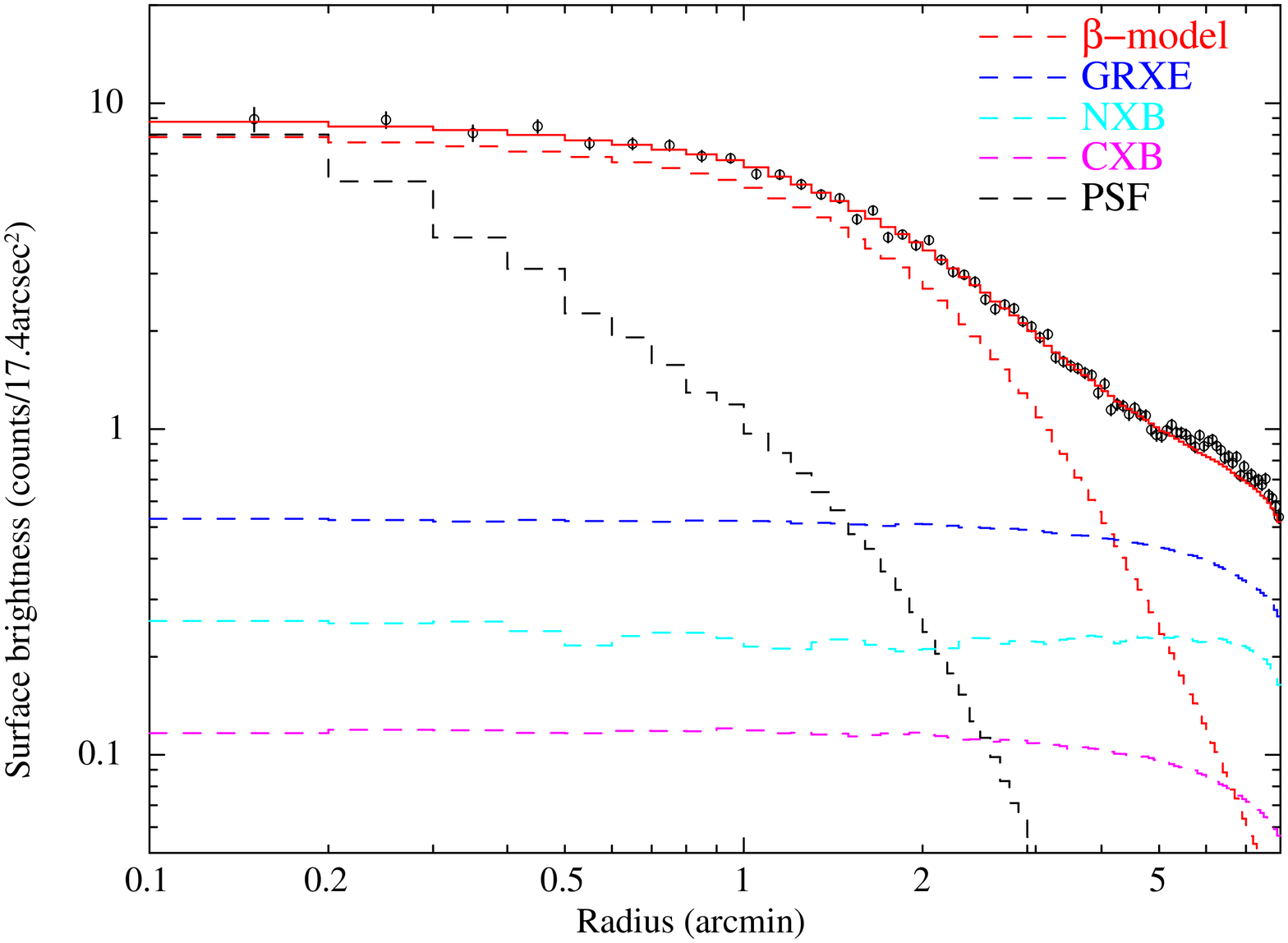}
 \end{center}
 \caption{Radial profile of the surface brightness of Suzaku
J1759$-$3450 in the $0.5$--$10$~keV band (open circles).  The error
bar on each data shows a $1 \sigma$ statistical uncertainty.  The
black dashed line represents the surface brightness profile of SS Cyg,
in which its brightness peak was normalized to that of Suzaku
J1759$-$3450.  The best-fit $\beta$-model with $(\beta, r_{\rm c}) =
(0.78, \timeform{1.61'})$, GRXE, NXB, and CXB components are shown
with red, blue, cyan, and magenta dashed lines, respectively.  The red
solid line indicates the combined model of these components.}
 \label{fig:radial_profile}
\end{figure}

To examine the spatial extent of the source quantitatively, we
generated a radial profile of the surface brightness from the XIS
image without the NXB subtraction and the vignetting correction, shown
in figure~\ref{fig:radial_profile}.  The origin of the radial profile
was set to be $(RA, Dec)_{\rm J2000.0} = (\timeform{17h59m17.41s},
\timeform{-34D50'18.6"})$, the position of the peak brightness.  The
surface brightness profile of SS Cyg (black dashed line in
figure~\ref{fig:radial_profile}), which corresponds to the point
spread function (PSF) of the XRTs, was plotted as well to illustrate
the diffuse emission.  We normalized the peak brightness of SS Cyg to
that of Suzaku J1759$-$3450.

We tried to reproduce the surface brightness profile of Suzaku
J1759$-$3450, $S(r)$, with a single $\beta$-model
\citep{1976A&A....49..137C}.  The $\beta$-model is frequently applied
to a self-gravitational system and is given by $S(r) \propto
(1+(r/r_{\rm c})^{2})^{-3 \beta + 1/2}$, where $r_{\rm c}$ and $\beta$
represent a core radius and a power-law index, respectively.  Because
of the point-like source located in the eastern direction of Suzaku
J1759$-$3450, we excised the outskirt of the radial profile ($r >
\timeform{5'}$) from the fit.

Since the XRT PSF and the vignetting effect are convolved in the
radial profile, we evaluated $\beta$ and $r_{\rm c}$ as described
below.  First, we created a sky image which surface brightness is
given by a $\beta$-model of $(\beta, r_{\rm c})$.  The peak brightness
was set to be at $(RA, Dec)_{\rm J2000.0} = (\timeform{17h59m17.41s},
\timeform{-34D50'18.6"})$.  Assuming that the X-ray emitting gas is an
isothermal optically-thin plasma with the temperature of $kT_{\rm e}
\sim 5$~keV and that the X-ray emission is modified with a foreground
absorption of $N_{\rm H} = 2.3 \times 10^{21}$~cm$^{-2}$, we simulated
the XIS events using \texttt{xissim}.  The exposure time and photon
flux were set to be large enough to obtain sufficient photon
statistics.  Using the simulated event list, we again made an X-ray
image and a radial profile of the surface brightness ($S_{\beta}(r)$)
in the same manner described above.  Then, the peak brightness was
normalized to be unity.

The X-ray emission from Suzaku J1759$-$3450 is contaminated by some
celestial and instrumental background components: GRXE, Cosmic X-ray
Background (CXB), and NXB.  The contribution of the NXB, $S_{\rm NXB}
(r)$, was estimated by generating the surface brightness profile from
the NXB image.  In order to estimate the CXB and GRXE contributions,
denoted by $S_{\rm CXB}(r)$ and $S_{\rm GRXE}(r)$, respectively, we
again simulated the XIS events assuming a uniform sky with a radius of
$\timeform{20'}$.  The procedure for deriving $S_{\rm CXB}(r)$ or
$S_{\rm GRXE}(r)$ was the same as that of $S_{\beta}(r)$.  The CXB
spectrum was assumed to be a power-law model with a cutoff energy of
$40$~keV \citep{1987IAUS..124..611B}.  The exposure time was set to be
the same as that of the observation.

On the other hand, for the GRXE component, we extracted the X-ray
spectrum from the outside of the background region (dashed ellipse in
figure~\ref{fig:Xray_images}a).  To remove the contribution of the
eastern point-like source, a circle with a radius of \timeform{2'}
centered on the source was excluded from the extraction region.  After
the NXB subtraction, we fitted the spectrum with a phenomenological
model: an optically-thin thermal plasma model with two different
temperatures, referred to as \citet{2013PASJ...65...19U}, plus the CXB
model described above.  The absorption column density was fixed to be
$N_{\rm H} = 2.3 \times 10^{21}$~cm$^{-2}$ again.  The best-fit
parameters of the plasma temperatures were $0.24$~keV and $7.4$~keV.
\citet{2013PASJ...65...19U} also derived the ($l$, $b$)-dependence of
the GRXE intensity.  The X-ray flux of the plasma component in the
$2.3$--$8$~keV band was $7.2 \times
10^{-4}$~photons~s$^{-1}$~cm$^{-2}$, consistent with that estimated
from the equation (1) in \citet{2013PASJ...65...19U} ($7.0 \times
10^{-4}$~photons~s$^{-1}$~cm$^{-2}$).  Hence, our evaluation of the
GRXE spectral model and its photon flux was verified.  The exposure
time in the GRXE simulation was set to be $2000$ times larger than
that of the observation to reduce the statistical fluctuation.

\begin{figure}[htp]
 \begin{center}
  \FigureFile(80mm,50mm){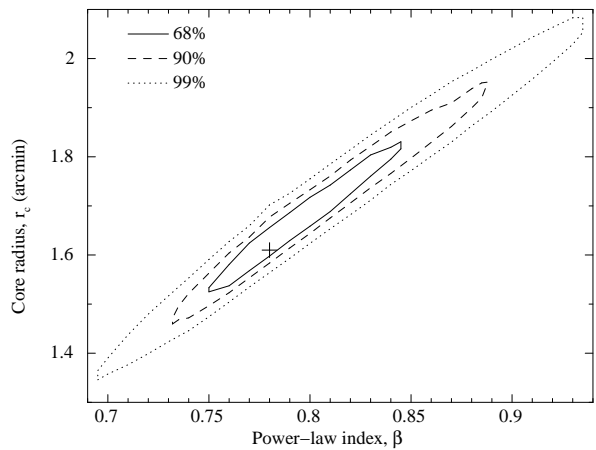}
 \end{center}
 \caption{Confidence contours of $\beta$ and the core radius.  The
solid, dashed, and dotted lines represent the $68$\% ($\Delta \chi
^{2} = 2.30$), $90$\% ($\Delta \chi ^{2} = 4.61$), and $99$\% ($\Delta
\chi^{2} = 9.21$) confidence levels, respectively.  The parameter set
of $(\beta, r_{\rm c}) = (0.78, \timeform{1.61'})$ we adopted here is
shown by a cross.}
 \label{fig:confidence_contours}
\end{figure}

We fitted the radial profile of Suzaku J1759$-$3450 in the range of
$\timeform{0.1'}$--$\timeform{5.0'}$ with a combined model of these
components as follows: $S(r) = a \times S_{\beta}(r) + S_{\rm GRXE}(r)
+ S_{\rm CXB}(r) + S_{\rm NXB}(r)$.  Here $a$ represents the
normalization of the $\beta$-model component.  For a given $(\beta,
r_{\rm c})$, the $\chi ^{2}$ value was calculated from the best-fit
parameters of $a$.  We examined the appropriate $\beta$ and $r_{\rm
c}$ which yield the $\chi ^{2}$ minimum by searching the parameter
space constructed from a set of $(\beta, r_{\rm c})$.  Since the
$\beta$ and $r_{\rm c}$ were strongly correlated with each other, the
unique determination of $(\beta, r_{\rm c})$ was difficult.
Therefore, we adopted the parameter combination of $(\beta, r_{\rm c})
= (0.78, \timeform{1.61'})$, which gave an acceptable fit of $\chi
^{2} = 58/50$~d.o.f..  The confidence contours of $(\beta, r_{\rm c})$
is shown in figure~\ref{fig:confidence_contours}.

\subsection{Spectral analysis}
\label{subsection:spectral_analysis}

\begin{figure*}[htbp]
 \begin{center}
  \FigureFile(80mm,50mm){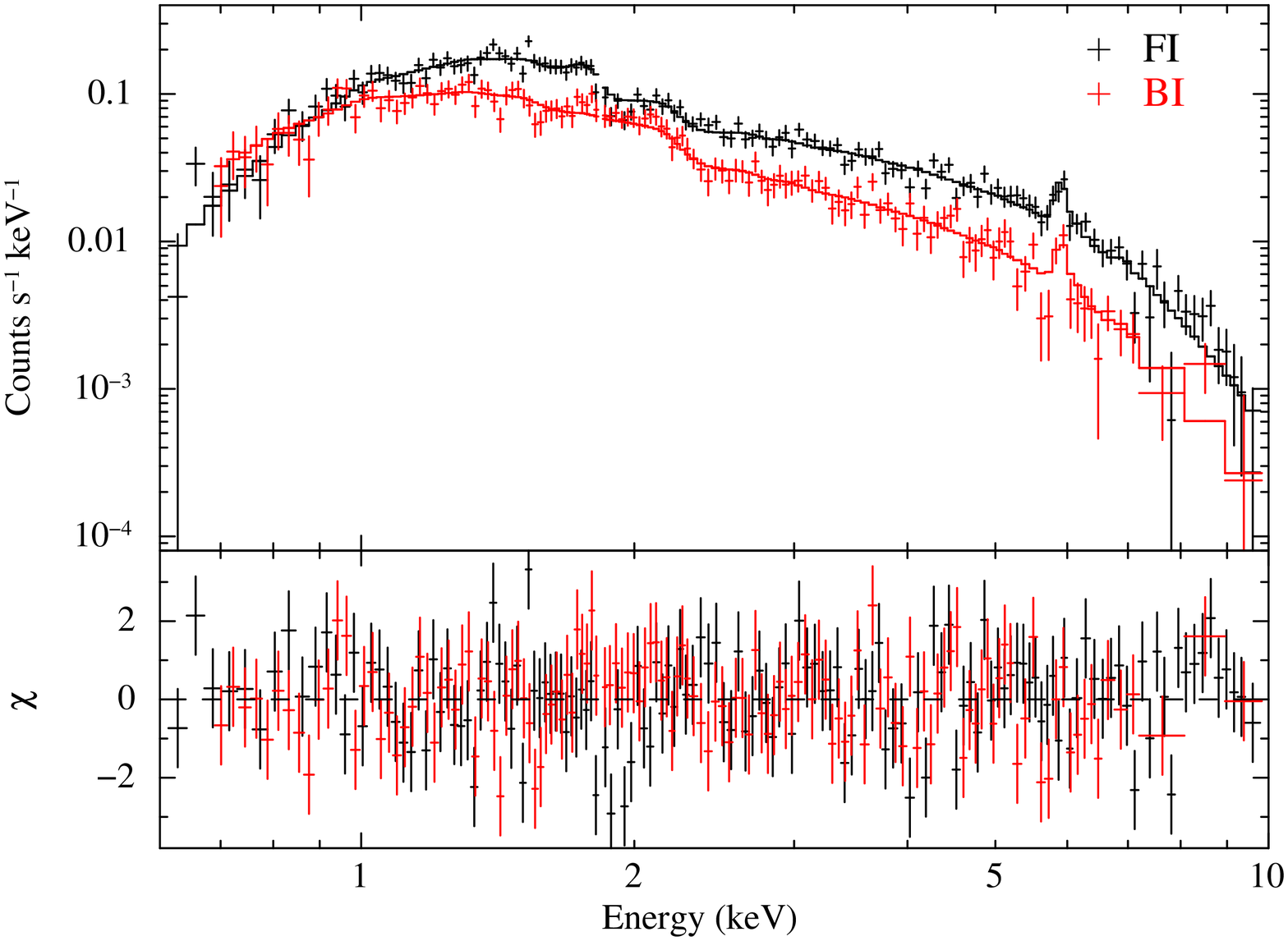}
  \FigureFile(80mm,50mm){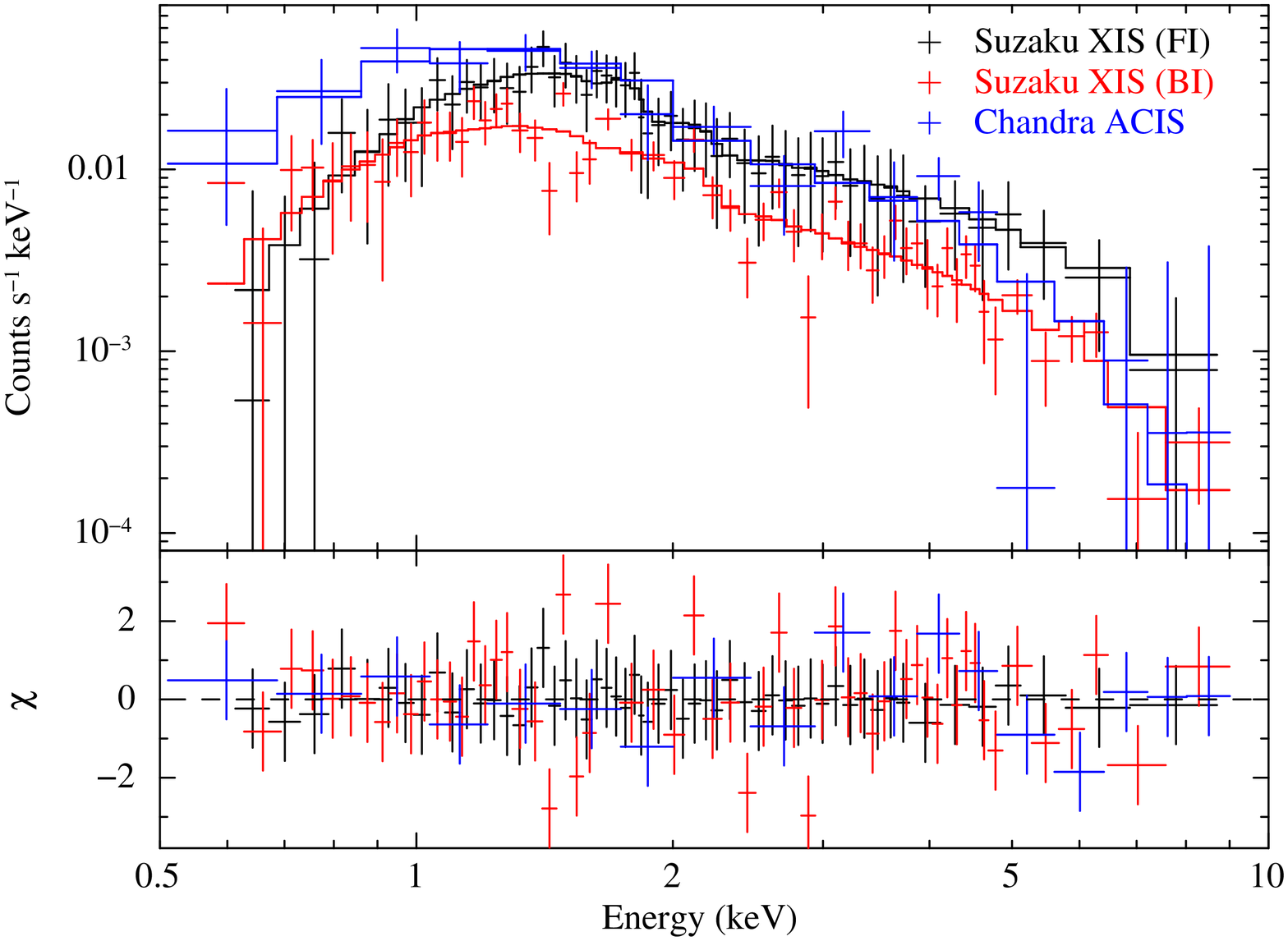}
 \end{center}
 \caption{X-ray spectra of Suzaku J1759$-$3450 obtained with the
Suzaku XIS (a) and the Chandra ACIS-S3 (b).  In the panel (b), the XIS
spectra extracted from the Chandra source region (see
figure~\ref{fig:Xray_images}b) are also shown.  The spectra of the FI
and BI CCDs are represented with black and red pluses, while the
Chandra spectrum is done with blue pluses.  In order to increase the
photon statistics of the FI spectrum, we took an average of the XIS0
and XIS3 spectra.  The best-fit optically-thin thermal plasma models
are overlaid with solid curves.}
 \label{fig:XIS_spectra}
\end{figure*}

We extracted the source and background spectra for each CCD sensor
from the respective regions shown in figure~\ref{fig:Xray_images}a.
We made the averaged spectrum for the FI CCDs.  The ARF was made by
\texttt{xissimarfgen} assuming the X-ray surface brightness derived in
the previous section: a single $\beta$-model with $(\beta, r_{\rm c})
= (0.78, \timeform{1.61'})$.  For the XIS1, the 2nd trailing rows were
excluded from the ARF calculation.  Figure~\ref{fig:XIS_spectra}a
shows the XIS spectra.

We found a conspicuous line-like emission at $\sim 6$~keV.  Although
there were some subtle structures over the whole energy band, we first
tried to fit the spectra with a power-law continuum plus Gaussian
emission line attenuated by photoelectric absorption
\citep{1992ApJ...400..699B}.  The $\sigma$ of the Gaussian function
was fixed to be $0$.  Taking into the consideration the uncertainty of
the relative normalization between FI and BI CCDs, we multiplied the
spectra by a constant factor; the factor of the FI spectrum was fixed
to be unity and that of the BI spectrum was free parameter.  We
excluded the $1.82$--$1.84$~keV band from the spectral fit due to the
uncertainty of the quantum efficiency around the Si edge.

This model yielded a marginally acceptable fit with $\Gamma = 2.1$ and
$N_{\rm H} = 3.8 \times 10^{21}$~cm$^{-2}$ ($\chi ^{2} =
346/273$~d.o.f.).  Thus, we introduced an exponential cutoff in the
power-law continuum.  The cutoff energy of $\sim 5$~keV improved the
fit significantly ($\chi ^{2} = 315/272$~d.o.f.).  The presence of the
exponential cutoff implies that the continuum emission has a thermal
origin.  We then replaced the cutoff power-law model with a thermal
bremsstrahlung.  The absorbed bremsstrahlung model including the
Gaussian emission line gave again an acceptable fit of $\chi ^{2} =
315/273$~d.o.f..  The electron temperature and the absorption column
density were $kT_{\rm e} = 5.0^{+0.4}_{-0.3}$~keV and $N_{\rm H} =
(2.1 \pm 0.3) \times 10^{21}$~cm$^{-2}$, respectively.  The center
energy of the emission line was found to be
$5.91^{+0.02}_{-0.03}$~keV.

The emission line at $\sim 6$~keV indicates that the X-ray emitting
gas is not in the local universe.  Hence, we tried to fit the spectra
by absorbed thermal emission from an optically-thin plasma in
collisional ionization equilibrium (\texttt{APEC} in \texttt{XSPEC})
with the redshift parameter thawed.  The elemental abundance relative
to the solar value \citep{1989GeCoA..53..197A} was also set to be free
parameter.  Again, we obtained an acceptable fit ($\chi ^{2} =
311/273$~d.o.f.).  The best-fit model and parameters are shown in
figure~\ref{fig:XIS_spectra}a and
table~\ref{table:best-fit_parameters}, respectively.  The absorption
column density of $N_{\rm H} = (2.3 \pm 0.3) \times 10^{21}$~cm$^{-2}$
and the electron temperature of $kT_{\rm e} = 5.3^{+0.5}_{-0.3}$~keV
were consistent with those obtained by the bremsstrahlung model.  The
elemental abundance was determined to be $Z/Z_{\odot} = 0.18 \pm
0.05$.  The redshift parameter was found to be $z =
0.132^{+0.005}_{-0.003}$, implying that the center energy of the
emission line at the rest frame is $6.7$~keV.  Thus, the emission line
is identified with K$\alpha$ emission from helium-like Fe ions
(Fe\emissiontype{XXV}).  We note that the spectral model and its
best-fit parameters were consistent with those assumed in the analysis
of the surface brightness profile (see
section~\ref{subsection:image_lightcurve}).  Hence, the
self-consistency for the estimation of $(\beta, r_{\rm c})$ was
verified.

\begin{table}
\begin{center}
\caption{Best-fit parameters of the Suzaku J1759$-$3450 spectra\footnotemark
[$*$].}
\label{table:best-fit_parameters}
\begin{tabular}{lll}
\hline
Parameters  & Source & Chandra region \\
\hline
$N_{\rm H}$  ($10^{21}$~cm$^{-2}$) & $2.3 \pm 0.3$ & $2.3$ (fixed) \\
$kT_{\rm e}$ (keV) & $5.3^{+0.5}_{-0.3}$ & $6.3^{+1.5}_{-1.2}$ \\
$Z/Z_{\odot}$      & $0.18 \pm 0.05$ & $0.18$ (fixed) \\
Redshift $z$      & $0.132^{+0.005}_{-0.003}$ & $0.132$ (fixed) \\
\hline
Flux (erg~s$^{-1}$~cm$^{-2}$)\footnotemark[$\dagger$] & $4.6 \times 10^{-12}$ & --- \\
$\chi ^{2}$/d.o.f. & 311/273 = 1.1 & 104/128 = 0.8 \\
\hline
\multicolumn{2}{@{}l@{}}{\hbox to 0pt{\parbox{80mm} {\footnotesize
 \footnotemark[$*$] Superscript and subscript figures represent the
upper and lower limit of the 90\% confidence interval, respectively.
 \par\noindent
 \footnotemark[$\dagger$] In the $0.5$--$10$~keV band.
}\hss}}
\end{tabular}
\end{center}
\end{table}

\subsubsection{Chandra spectrum}
\label{subsubsection:chandra_spectrum}
The Chandra observation covered only a small fraction of the Suzaku
FOV since we applied the window option with a width of $128$~pixel.
Nevertheless, we found the X-ray emission from Suzaku J1759$-$3450
within the Chandra FOV. (see figure~\ref{fig:Xray_images}b).  Then, we
extracted the spectrum from the source region where the Chandra FOV
and the source region of the XIS spectra (see
section~\ref{subsection:image_lightcurve}) overlap.  The
background spectra was also generated from the overlapping background
region, defined in the same manner as the source region.  The
background-subtracted spectrum was grouped to contain at least
$15$~photons per bin.

We show the Chandra spectrum of Suzaku J1759$-$3450 in
figure~\ref{fig:XIS_spectra}b.  Since the net count of the spectrum is
only $217$~photons, the spectrum did not provide sufficient
constraints on the spectral parameters of the diffuse emission.
Hence, we created the XIS spectra from this source region shown by
black and red pluses in figure~\ref{fig:XIS_spectra}b, and then
performed a joint fit to the Chandra and Suzaku spectra.  For the XIS
spectra, the background spectra used in the previous section were
subtracted to minimize the statistical uncertainty.

We fitted these spectra with the thermal plasma model (\texttt{APEC}
in \texttt{XSPEC}) multiplied by photoelectric absorption.  In this
fit, we fixed the parameters of the absorption column density ($N_{\rm
H}$), the elemental abundance, and redshift ($z$) to be those derived
from the Suzaku spectra; the electron temperature and the
normalization were free parameters.  The absorbed thermal plasma model
gave an acceptable fit ($\chi ^{2} = 104/128$~d.o.f.).  The best-fit
electron temperature was $6.3^{+1.5}_{-1.2}$~keV (see also
table~\ref{table:best-fit_parameters}).  The temperature was
consistent with that of the X-ray emission extracted from the Suzaku
source region.

\section{Discussion}
\label{section:discussion}
We conducted the Chandra and Suzaku observations of 1RXS
J175911.0$-$344921 to perform the imaging spectroscopy above $2$~keV
for the first time.  No point-like X-ray counterpart of 1RXS
J175911.0$-$344921 was found both in the Chandra and Suzaku FOVs.  On
the other hand, we found diffuse X-ray emission with a nearly circular
shape, designated as Suzaku J1759$-$3450.  The source extent of
$\timeform{63"}$ for 1RXS J175911.0$-$344921 indicates extended X-ray
emission compared with the ROSAT PSF towards this RBSC source.  Thus,
the large spatial extent may cause a false determination of the source
position.  The surface brightness profile and the extragalactic origin
of the diffuse X-ray emission indicate that Suzaku J1759$-$3450 is a
new cluster of galaxies.

The X-ray profile of the surface brightness in the $0.5$--$10$~keV
band did not show a strong central peak and was reproduced with a
single $\beta$-model.  The core radius and the power-law index were
estimated to be $r_{\rm c} = \timeform{1.61'}$ and $\beta = 0.78$,
respectively.  The power-law index we adopted here is in the range of
the $\beta$ parameters obtained by \citet{2004A&A...428..757O}.  We
note that the peak position of the surface brightness at $(RA,
Dec)_{\rm J2000.0} = (\timeform{17h59m17.41s},
\timeform{-34D50'18.6"}$) was slightly shifted in the south-east
direction from the center of the circular emission whose spectrum we
extracted.  The non-axisymmetric morphology suggests that the X-ray
emitting gas is far from relaxed.  No central cooling core also
suggests that the hot gas is more likely a merging system.

The X-ray spectrum of Suzaku J1759$-$3450 showed a remarkable emission
line at $\sim 6$~keV.  Taking into account the cosmological redshift,
we reproduced the spectrum by an optically-thin thermal plasma with an
electron temperature of $kT_{\rm e} = 5.3$~keV.  Since the redshift
parameter was determined to be $z = 0.132$, the origin of the emission
line was identified to be highly-ionized Fe.  The absorption column
density of $N_{\rm H} = (2.3 \pm 0.3) \times 10^{21}$~cm$^{-2}$ was
consistent with the H\emissiontype{I} column density of $2.4 \times
10^{21}$~cm$^{-2}$ \citep{1990ARA&A..28..215D}.  This result
reinforces that Suzaku J1759$-$3450 is an extragalactic object
suffered from the local ($z = 0$) absorption due to the Galactic
interstellar medium.  The redshift parameter of $z = 0.132$
corresponds to the distance to the source of $550$~Mpc.  The angular
separation of $\timeform{1'}$ on the sky is equivalent to the distance
of $140$~kpc.  Thus, the core radius of Suzaku J1759$-$3450 is $\sim
230$~kpc.

Assuming that the hot gas of Suzaku J1759$-$3450 is in hydrostatic
equilibrium and that its density profile follows an isothermal
$\beta$-model, we can estimate the total mass of the cluster within a
given radius as follows (e.g., \cite{1980ApJ...241..552F}):
\begin{equation}
M (< r) = \frac{3 \beta k T_{\rm e} r^{3}}{G \mu m_{\rm H} r^{2}_{\rm c}} \frac{1}{1 + (r/r_{\rm c})^2}.
\end{equation}
Here, $k$, $G$, $\mu = 0.6$, and $m_{\rm H}$ represent the Boltzmann
constant, the gravitational constant, the mean molecular weight, and
the mass of the hydrogen, respectively.  Although we assumed here that
the gas was isothermal, no temperature structure was indeed found in
the joint analysis of the Chandra and Suzaku spectra.  Using the
$(\beta, r_{\rm c}) = (0.78, \timeform{1.61'})$, we estimated the
total mass within $r < 4'$ to be $2.2 \times 10^{14} M_{\odot}$,
consistent with that of a typical cluster of galaxies.

The X-ray flux in the $0.5$--$10$~keV band of $4.6 \times
10^{-12}$~erg~s$^{-1}$~cm$^{-2}$ is converted into the luminosity of
$2.1 \times 10^{44}$~erg~s$^{-1}$ at $z = 0.132$.  The unabsorbed
bolometric luminosity is also estimated to be $L_{\rm X} (r < 4') =
3.5 \times 10^{44}$~erg~s$^{-1}$.  In order to compare our result with
the $L_{\rm X}$--$T$ relation which has been ever studied, we
calculated the overdensity radius ($r_{\rm 500}$) within which the
averaged density ($\bar{\rho}$) of the cluster is $500$ times larger
than the critical density of the universe ($\rho _{\rm crit} = 9.2
\times 10^{-30}$~g~cm$^{-3}$).  Since the averaged density can be
evaluated by
\begin{equation}
\bar{\rho} (<r) = \frac{9 \beta k T_{\rm e}}{4 \pi G \mu m_{\rm H}
r^{2}_{\rm c}} \frac{1}{1 + (r/r_{\rm c})^{2}},
\end{equation}
we estimated to be $r_{500} = 5.5 r_{\rm c} = \timeform{8.9'}$, which
corresponds to the distance of $1.3$~Mpc.  A new ARF was re-calculated
with \texttt{xissimarfgen} for the source region with a radius of
$\timeform{8.9'}$.  Using the best-fit parameters derived from the
spectral analysis, the resultant bolometric luminosity was estimated
to be $L_{\rm X} (r < r_{\rm 500}) = 4.3 \times 10^{44}$~erg~s$^{-1}$.

We also investigated the systematic uncertainty of the X-ray
luminosity due to the $\beta$-model assumed in the ARF calculation.
The two cases were taken into consideration based on the $90$\%
confidence contour shown in figure~\ref{fig:confidence_contours}:
$(\beta, r_{\rm c}) = (0.732, \timeform{1.460'})$ (lower limit) and
$(0.888, \timeform{1.952'})$ (upper limit).  The XIS spectra were
re-fitted with the optically-thin thermal plasma model using the ARFs
assumed in these cases.  The best-fit parameters of $N_{\rm H}$,
$kT_{\rm e}$, the abundance, and the redshift $z$ were all consistent
with those derived from the $(\beta, r_{\rm c}) = (0.78,
\timeform{1.61'})$ case.  The bolometric luminosity changed by $2$\%:
$L_{\rm X} (r < r_{\rm 500}) = 4.2 \times 10^{44}$~erg~s$^{-1}$ in the
lower-limit case and $4.4 \times 10^{44}$~erg~s$^{-1}$ in the
upper-limit case.

According to the equation (5) in \citet{2006ApJ...640..673O}, the
expected bolometric luminosity at $kT_{\rm e} = 5.3$~keV is $1.2
\times 10^{45}$~erg~s$^{-1}$, 3 times larger than that of Suzaku
J1759$-$3450.  Taking into consideration a relatively large scatter in
the $L_{\rm X}$--$T$ relation shown in figure 5a of
\citet{2006ApJ...640..673O}, however, we should conclude that the
combination of the temperature and bolometric luminosity within
$r_{500}$ of Suzaku J1759$-$3450 is consistent with that of a typical
cluster.  Suzaku J1759$-$3450 may be classified into relatively dim
cluster of galaxies based on its luminosity.
\citet{2012MNRAS.421.1583M} argued that unrelaxed clusters of galaxies
with temperature below $6$~keV tend to show the X-ray luminosities
lower than those of the relaxed ones.  Our conjecture that Suzaku
J1759$-$3450 is not a relaxed system supports their claim.  We note
that the overdensity radius of $r_{500} \sim 1$~Mpc also agrees with a
hot gas associated with a cluster of galaxy, since the overdensity
radius depends weakly on the redshift parameter and concentrates
around $1$~Mpc (e.g., \cite{2002ApJ...567L..23O}).

\section{Summary}
\label{section:summary}
We discovered extended hard X-ray emission above $2$~keV towards the
unidentified X-ray source, 1RXS J175911.0$-$344921, with the Suzaku
and Chandra observations.  Thanks to the low background level,
characteristic of the Suzaku's performance, the spectrum of this
diffuse X-ray emission was obtained with sufficient photon statistics.
We summarize the image and spectral analyses of this diffuse emission,
designated as Suzaku J1759$-$3450, as below.

\begin{enumerate}
\item The radial profile of the surface brightness of Suzaku
J1759$-$3450 was explained with the isothermal $\beta$-model.  The
core radius and power-law index were estimated to be $r_{\rm c} =
\timeform{1.61'}$ and $\beta = 0.78$, respectively.
\item The Suzaku J1759$-$3450 spectrum was well reproduced with the
X-ray emission from an optically-thin thermal plasma with the
temperature of $kT_{\rm e} = 5.3^{+0.5}_{-0.3}$~keV, modified with the
Galactic absorption of $N_{\rm H} = (2.3 \pm 0.3) \times
10^{21}$~cm$^{-2}$.  The emission line at $\sim 6$~keV was identified
with the redshifted ($z = 0.132$) K$\alpha$ emission from the
helium-like Fe ions.  The elemental abundance was $0.18 \pm 0.05$
relative to the solar value.
\item Assuming that the X-ray emitting gas is an isothermal sphere in
hydrostatic equilibrium, the total mass of this cluster of galaxies
was estimated to be $2.2 \times 10^{14} M_{\odot}$.  The unabsorbed
bolometric luminosity within $r_{500}$ of $4.3 \times
10^{44}$~erg~s$^{-1}$ was slightly lower than that expected from the
electron temperature of $kT_{\rm e} = 5.3$~keV and the $L_{\rm
X}$--$T$ relation.  However, the relatively large scatter in the
$L_{\rm X}$--$T$ relation allows Suzaku J1759$-$3450 to be considered
as a new member of cluster of galaxies.
\end{enumerate}

\bigskip
We appreciate the helpful comments from an anonymous referee to
improve our manuscript.  We would like to thank all the Suzaku team
members for their support of the observation and useful information on
the XIS and HXD analyses.  We are also grateful to Prof. Tawara for
his useful comments on the spectral analysis.

\end{document}